\documentclass[aps,prb,twocolumn,amsmath,amssymb,superscriptaddress]{revtex4-1}
\usepackage{amsmath}
\usepackage{amssymb}
\usepackage{amsthm}
\usepackage{graphicx}
\usepackage{dcolumn} 
\usepackage{bm} 
\usepackage{epic}
\usepackage{bbm}
\usepackage{longtable}
\usepackage{slashed}
\usepackage{ulem}   
\begin{document}

\title{
Coexistence of Fermi arcs with two dimensional gapless Dirac states}

\author{Adolfo G.\ Grushin} 
\affiliation{Max-Planck-Institut f\"ur Physik komplexer Systeme, 01187
  Dresden, Germany}       
\author{J\"orn W.\ F.\ Venderbos} 
\affiliation{Department of Physics, Massachusetts Institute of
  Technology, Cambridge, Massachusetts 02139, USA}
\author{Jens H.\ Bardarson} 
\affiliation{Max-Planck-Institut f\"ur Physik komplexer Systeme, 01187
  Dresden, Germany}

\date{\today}

\begin{abstract}
We present a physical scenario in which both Fermi arcs and two-dimensional gapless Dirac states coexist as boundary modes at the \textit{same} two-dimensional surface. This situation is realized in topological insulator--Weyl semimetal interfaces in spite of explicit time reversal symmetry breaking. 
Based on a heuristic topological index, we predict that the coexistence is allowed when (i) the corresponding states of the Weyl semimetal and topological insulator occur at disconnected parts of the Brillouin zone separated by the Weyl nodes and (ii) the time-reversal breaking vector defining the Weyl semimetal has no projection parallel to the domain wall. This is corroborated by a numerical simulation of a tight binding model. We further calculate the optical conductivity of the coexisting interface states, which can be used to identify them through interference experiments. \end{abstract}

\maketitle

\emph{Introduction}---Protected surface states are the key
characteristic of topological phases of matter. 
Time reversal invariant topological insulators (TI) host at their surface two-dimensional (2D) massless Dirac quasiparticles protected by time reversal symmetry ($\mathcal{T}$)~\cite{HK10,QZ11,HM11}.
Weyl semimetals (WSM) are 3D gapless materials described at low energy by Weyl fermions. 
They host topologically robust surface states referred to as Fermi arcs since they form an open Fermi surface~\cite{Wan:2011hi,Burkov:2011de,WK12,Hosur:2013eb,Turner:2013tf,Vafek:2014hl}. 
Their emergence can be understood in terms of charge conservation (gauge invariance) of the effective field theory describing the WSM's response to external electromagnetic fields, in analogy with the quantum Hall effect \cite{Grushin:2012cb,Goswami:2013jp, Ramamurthy:2014uh}. 
Such an effective response predicts a number of striking physical properties, such as a finite Hall conductivity~\cite{Burkov:2011de,Aji:2012gs,Grushin:2012cb,Goswami:2013jp,Liu:2013kv}, a current parallel to an external magnetic field (chiral magnetic effect)~\cite{Zyuzin:2012ca,Grushin:2012cb,Goswami:2013jp,Landsteiner:2014fw}, and a finite angular momentum induced by a thermal gradient (axial magnetic effect) \cite{CCG14}.

The band structure of a WSM is characterized by a linear dispersion
around a set of non-degenerate band touchings points called Weyl
nodes.
Their existence requires the breaking of either time reversal $\mathcal{T}$ or inversion $\mathcal{I}$ symmetry~\cite{Turner:2013tf,Vafek:2014hl}. 
Each Weyl node is chiral and has an associated momentum space Berry flux that gives it the character of a Berry flux monopole.
Since the total flux in momentum space is required to be zero by gauge invariance~\cite{BH13}, the Weyl nodes must appear in pairs with opposite monopole charge~\cite{NielNino81a,NielNino81b}. 
They are therefore topologically stable as they can only be annihilated by bringing together a pair with opposite chirality~\cite{KLINKHAMER:2005bi}.
The simplest realization of a topological semimetal, one exhibiting only a single pair of nodes, necessarily breaks time-reversal symmetry.
Indeed, $\mathcal{T}$ symmetry connects two Weyl nodes with the \emph{same} monopole charge~\cite{YZT12}, implying the existence of at least another pair with opposite chirality~\cite{HB12,O13}.
From this symmetry perspective, the coexistence of 2D Dirac TI surface states and pairs of Fermi arcs is in principled allowed if $\mathcal{T}$ symmetry is respected, as inferred from ab-initio calculations~\cite{WSC12}.
However, in the minimal two-Weyl-node model, this symmetry is broken and such coexistence seems to be mutually exclusive; the 2D Dirac TI surface state is protected by $\mathcal{T}$ while Fermi arcs are only realized in its absence.
\begin{figure}
\includegraphics[scale=0.19]{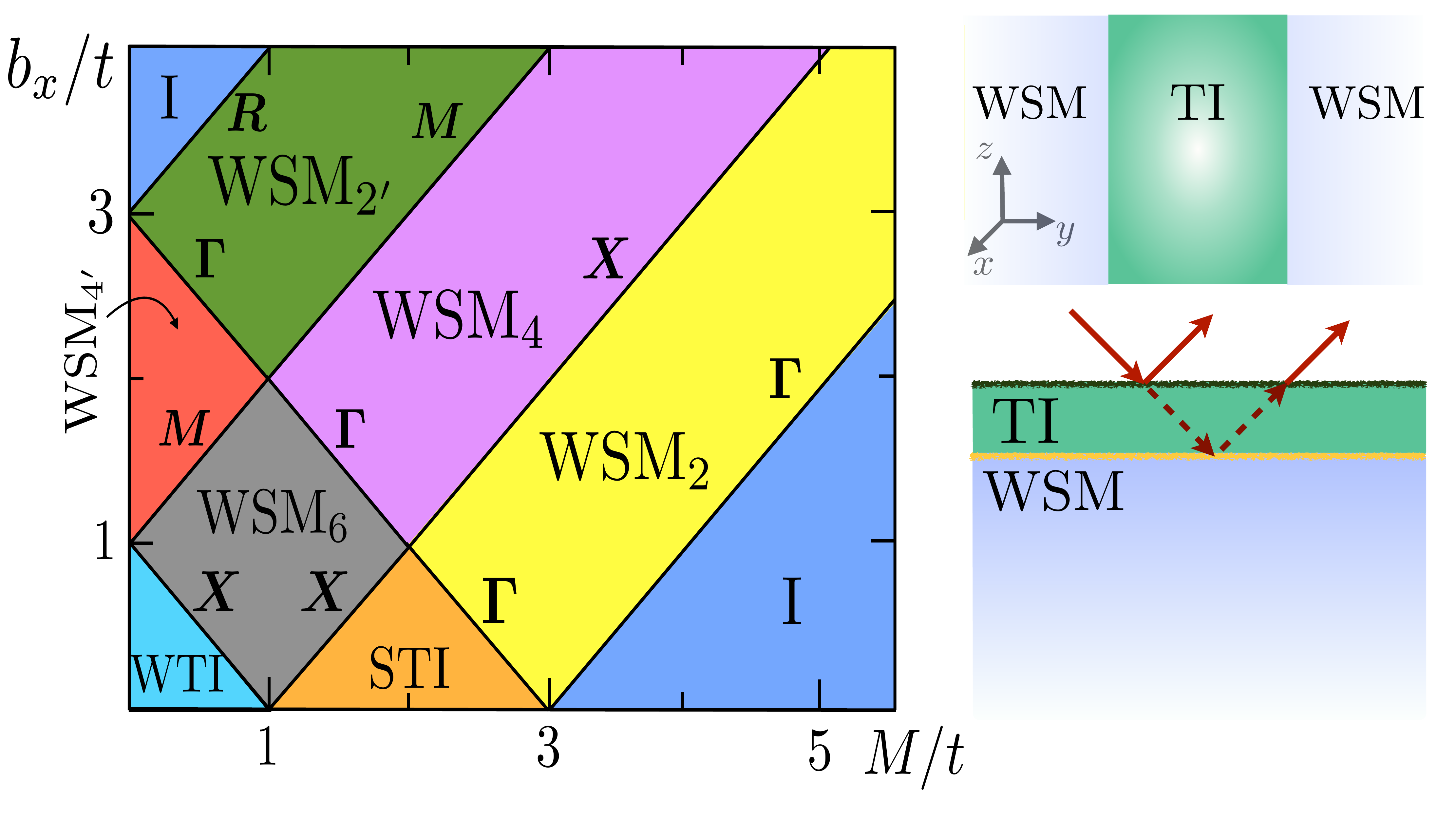}
\caption{\label{Fig:phasediag} (Color online) Left: Phase diagram for the model~\eqref{H} at $b_0 = 0$ as a function of $\mathbf{b}=(b_{x},0,0)$ and $M$. The four gapped phases are a weak (WTI) and a strong (STI) topological insulator and trivial (I) insulator. The Weyl semimetal phases have six (WSM$_{6}$), four (WSM$_{4,4'}$) or two Weyl nodes (WSM$_{2,2'}$). Solid lines represent gap closing or Weyl node annihilations at the corresponding Brillouin zone momenta. 
Upper right panel: Geometry, with periodic boundary conditions, used in numerical simulations.
Lower right panel: A schematic of a proposed interference experiment to observe the coexistence of surface states.}
\end{figure}

In this work we show how to circumvent this apparent dichotomy and realize both states at the \emph{same} surface, the interface of a WSM--TI heterostructure, and we calculate the optical conductivity of the coexisting interface states, which serves as their distinct experimental signature. 
We demonstrate this by modelling such interfaces using a canonical cubic lattice model describing TIs supplemented with symmetry breaking fields that can drive the system into a WSM phase~\cite{QHZ08,Vazifeh:2013fe}.
Spatially dependent parameter fields realize a generic model of a domain wall between two different phases. 
We numerically observe that coexistence occurs when the Fermi arcs and 2D massless Dirac
surface states occupy distinct parts of the Brillouin zone delimited by the Weyl nodes \footnote{By delimited it is to be understood that the Weyl nodes act like a gap closing topological phase transition for a 2D Hamiltonian defined by a cut in a direction perpendicular to the line connecting two Weyl nodes.}.
This observation is captured by a heuristic topological index
\begin{equation}\label{eq:index}
\mathcal{J}_{a}=C_{a}\pi_{a},
\end{equation}
defined for each surface time reversal invariant momenta $\Lambda_a$ and written in terms of known properties of the two phases (we use indices $a,b,\ldots$ to label surface momenta and $i,j,\ldots$ bulk momenta).
Namely, the time reversal polarizations $\pi_{a}=\pm 1$ determine the presence or absence of 2D Dirac surface states at $\Lambda_a$~\cite{FKM07,FK07} while $C_{a}=0\, (1)$ when a Fermi arc exists (is absent) in the vicinity of $\Lambda_a$.
By computing the index $\mathcal{J}_a$ for all $\Lambda_a$ one can predict whether coexistence of both surface states is allowed in a given surface or not (details are given below).

\emph{Coexistence of Fermi arcs and 2D massless Dirac states}---We model the bulk phases with the two orbital spinful cubic lattice Hamiltonian~\cite{QHZ08,Vazifeh:2013fe}
\begin{subequations}
	\label{H}
\begin{align}
H &= H_{\mathrm{TI}} + H_{b}, \\
H_{\mathrm{TI}}&=t\sum_{\bf{x}, \hat{\mathbf{j}}} c^{\dagger}_{\mathbf{x}}\frac{\Gamma_0 - i\Gamma_j}{2}
c_{\mathbf{x}+ \hat{\mathbf{j}}} + \text{h.c.} + M \sum_{\mathbf{x}}c^{\dagger}_{\mathbf{x}} \Gamma_0 c_{\mathbf{x}},\\
H_{b} &= \sum_{\mathbf{x},\mu}b_{\mu}~c^{\dagger}_\mathbf{x} \Gamma_{\mu}^{(b)} c_{\mathbf{x}}.
\end{align}
\end{subequations}
The position vector $\mathbf{x}$ runs over the sites of the cubic lattice, $\hat{\mathbf{j}}=\hat{x},\hat{y},\hat{z}$ is a unit vector in each cartesian direction, and $\mu=0,x,y,z$. 
The operator $c_{\mathbf{x}}=(c_{\mathbf{x}A\uparrow},c_{\mathbf{x}A\downarrow}, c_{\mathbf{x}B\uparrow},c_{\mathbf{x}B\downarrow})$ where 
$c_{\mathbf{x}\sigma s}$ annihilates an electron in orbital $\sigma=A,B$ at $\mathbf{x}$ with spin $s = \uparrow,\downarrow$. 
The gamma matrices are given by $\Gamma^{\mu}=(\sigma_{x}\otimes s_{0},\sigma_{z}\otimes s_{y},\sigma_{z}\otimes s_{x},\sigma_{y} \otimes s_0)$ and $\Gamma_{\mu}^{(b)}=(\sigma_{y}\otimes s_{z},\sigma_{x} \otimes s_{x},\sigma_{x} \otimes s_{y},\sigma_0 \otimes s_{z})$, where $s_j$ and $\sigma_j$ are Pauli matrices describing the spin and orbital degree of freedom respectively and $s_0,\sigma_0$ are the corresponding identity operators.

The first term $H_{\mathrm{TI}}$ generally models a gapped insulator, apart from special parameter values that correspond to phase transitions between different insulating phases.
At the time reversal symmetric momenta $\Lambda_i$ of the cubic lattice, $\boldsymbol{\Gamma}=(0,0,0),\mathbf{X}=\mathcal{P}[(\pi,0,0)],\mathbf{M}=\mathcal{P}[(\pi,\pi,0)]$ and $\mathbf{R}=(\pi,\pi,\pi)$ with $\mathcal{P}$ the permutation operator, the gap is given by $2m_{i}$ with $m_{\boldsymbol{\Gamma},\boldsymbol{R}}=M\mp 3t$ and $m_{\boldsymbol{X},\boldsymbol{M}}=M\mp t$.
Depending on the relative signs of the masses one obtains a strong (STI, $t<|M|<3t$), a weak (WTI, $|M|<t$), or a trivial insulator (I, $|M| > 3t$), cf.\ the horizontal axis in the phase diagram of Fig.~\ref{Fig:phasediag}.

The term $H_b$ is parameterized by the four-vector $b_{\mu}=(b_{0},\mathbf{b})$, where the pseudo-scalar $b_{0}$ breaks $\mathcal{I}$ and the pseudo-three-vector $\mathbf{b}$ breaks $\mathcal{T}$.
For now, we focus on the case of main interest $b_0 = 0$ and take $\mathbf{b}=b_{x}\hat{x}$ without loss of generality. %
We consider the effect of a nonzero $b_0$ later.
With increasing $b_x$ the gap closes at one (or more) of the bulk time reversal symmetric momenta at which point the bulk spectrum is characterized by a 3D Dirac cone.
Upon further increase of $b_x$ this Dirac cone splits into two 3D Weyl nodes and a WSM phase is obtained.
Depending on the number of gap closings one obtains a WSM phase with two, four, or six Weyl nodes.
A fully representative corner of the phase diagram is provided in Fig.~\ref{Fig:phasediag}.

To model an interface between two distinct phases we endow the parameters of the Hamiltonian~\eqref{H} with a position dependence.
We limit ourselves to sharp interfaces parallel to the $x-z$ plane~\footnote{We have studied numerically the effect of smoothing the domain wall as opposed to a sharp interface. The reported interface features survive, details of which will be reported elsewhere.} with an infinite lattice in the $x$ and $z$ directions and a finite width $L_y$ in the $y$ direction.
To avoid interfaces with the vacuum, we take periodic boundary conditions in the $y$ direction resulting in two domain walls, see Fig.~\ref{Fig:phasediag}.
Explicitly, 
\begin{equation}
	(b_x,M) = 
	\begin{cases}
		(b_1,M_1) & \text{if } \frac{L_y}{4} < y < \frac{3L_y}{4}, \\
		(b_2,M_2) & \text{otherwise}.
	\end{cases}
\end{equation} 
Depending on the values of $(b_1,M_1)$ and $(b_2,M_2)$, we obtain an
interface between any two phases of the model; In the following we
focus on interfaces between a WSM and either a strong
or a weak TI. 

\begin{figure}
\includegraphics[width=\columnwidth]{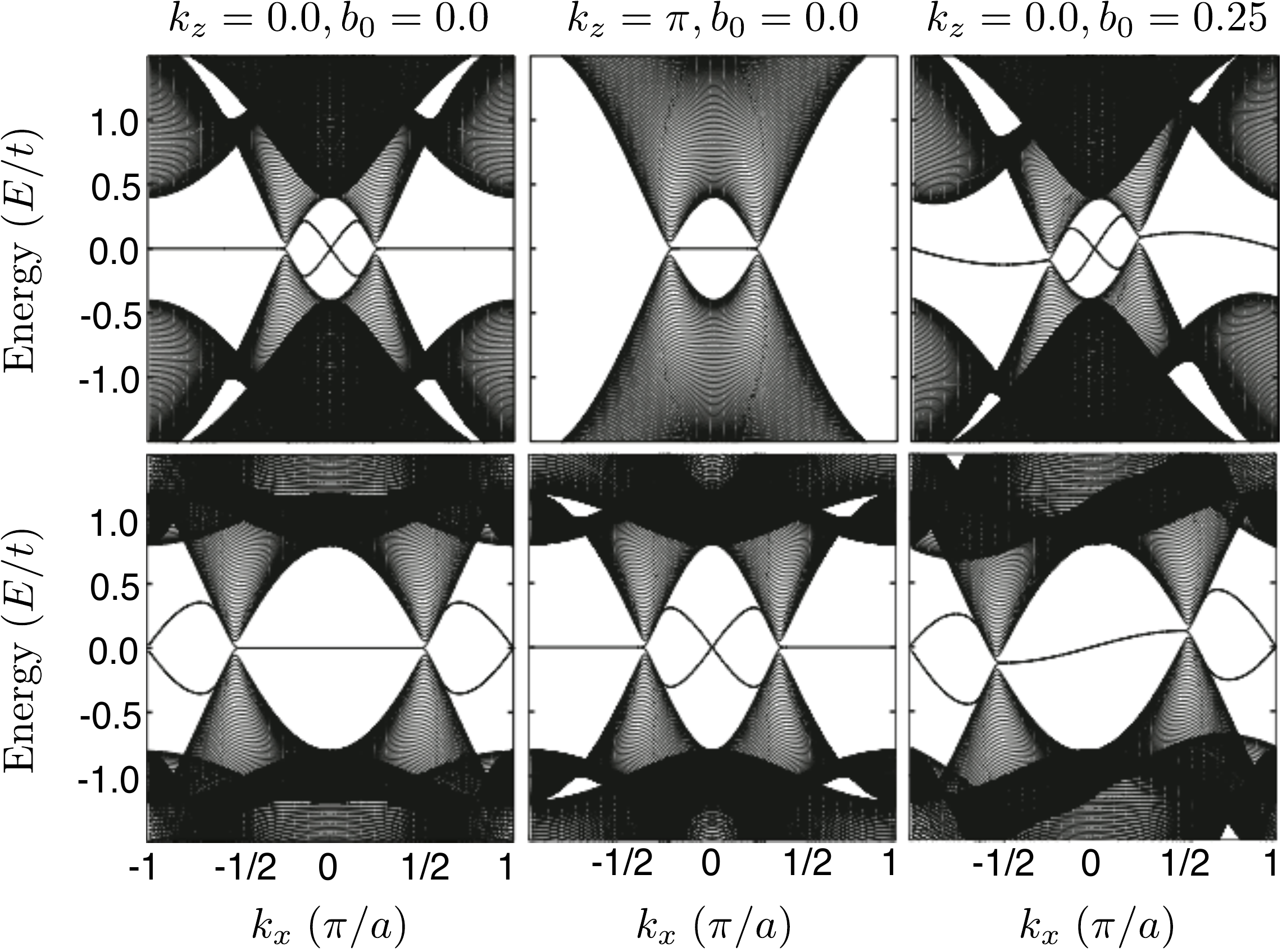}
\caption{\label{Fig:interpol} (Color online) Upper row: Band structure
  of a finite slab of WSM$_4$--STI heterostructure with periodic boundary
  conditions [$(b_x,M)_\text{STI} = (0,2.6)$
  and $(b_x,M)_{\text{WSM}_4} = (2,2.6)$]. Lower row: Band structure for a finite slab
  corresponding to a WSM$_{4'}$--WTI configuration [$(b_x,M)_{\text{STI}} = (0,0.2)$  and $(b_x,M)_{\text{WSM}_{4'}} = (2,0.2)$].  The
  $k_{z}=0,\pi$ cuts shown in the first and second columns demonstrate the
  coexistence of 2D Dirac states with Fermi arcs in both cases. The
  third column shows the effect of $b_{0}=0.25$. Band structures are
  obtained for systems with linear dimension $L_y = 160$.}
\end{figure}

In Fig.~\ref{Fig:interpol} we plot cuts through the energy spectrum for $k_{z}=0$ and $k_{z}=\pi$ as a function of $k_{x}$ for domain walls STI--WSM$_4$ and WTI--WSM$_{4'}$, demonstrating the coexistence at zero energy of Fermi arcs and massless 2D Dirac states separated by the Weyl nodes.
For the WTI and STI case there are an even and odd number of massless Dirac states at the interface respectively.
The zero modes are doubly degenerate with one state localized at each domain wall.
Importantly, not all STI--WSM or WTI--WSM domain walls have coexisting Fermi arc and Dirac states.
For instance, at a STI--WSM$_{4'}$ interface (data not shown), where both types of surface states would have to exist in the same region of momentum space, only Fermi arcs are numerically observed. 
This suggests that coexistence at the interface is allowed only as long as the states do not overlap in momentum space.

To assess the robustness of these states, we have studied the effect of a finite $\mathcal{I}$ breaking term $b_{0}$ in the WSM side of the domain wall. 
For a bulk WSM nonzero $b_{0}$ acts as a relative energy shift for the Weyl nodes and may lead to the chiral magnetic effect, the existence of which is still debated~\cite{Volovik:1999da,Aji:2012gs,Zyuzin:2012ca,Grushin:2012cb,Zyuzin:2012kl,Goswami:2013jp,Liu:2013kv,Vazifeh:2013fe,Chen:2013ep,Goswami:wl,Hosur:2012fl,Landsteiner:2014fw}.
At the interface $b_0$ endows the Fermi arcs with finite dispersion and tilts the 2D Dirac states, see right column of Fig.~\ref{Fig:interpol}.
Increasing $b_{0}$ further ultimately opens up a gap in the WSM and destroys the Fermi arcs~\cite{Zyuzin:2012ca,Grushin:2012cb}. 

The coexistence is, furthermore, only allowed when $\mathbf{b}$ is perpendicular to the domain wall direction $\hat{y}$; a finite parallel component ($b_y\neq 0$) acts as a Zeeman term for the 2D Dirac surface states and opens up a gap (as we have verified numerically).
This effect is minimized by aligning the domain wall along $\mathbf{b}$, which physically is an intrinsic magnetization and likely to be aligned with an experimentally identifiable crystallographic direction. 

\emph{Coexistence from bulk topology---}Our numerical results are captured by the topological index $\mathcal{J}_{a}$ defined in~Eq.~\eqref{eq:index}, the construction and use of which we now explain. 
The index relies on the observation that the bulk properties of the WSM and TI impose conditions on where their corresponding surface states must occur.
When they are all compatible, the coexistence is allowed and protected by the presence of the Weyl nodes.

\begin{figure}
\includegraphics[scale=0.19,page=1]{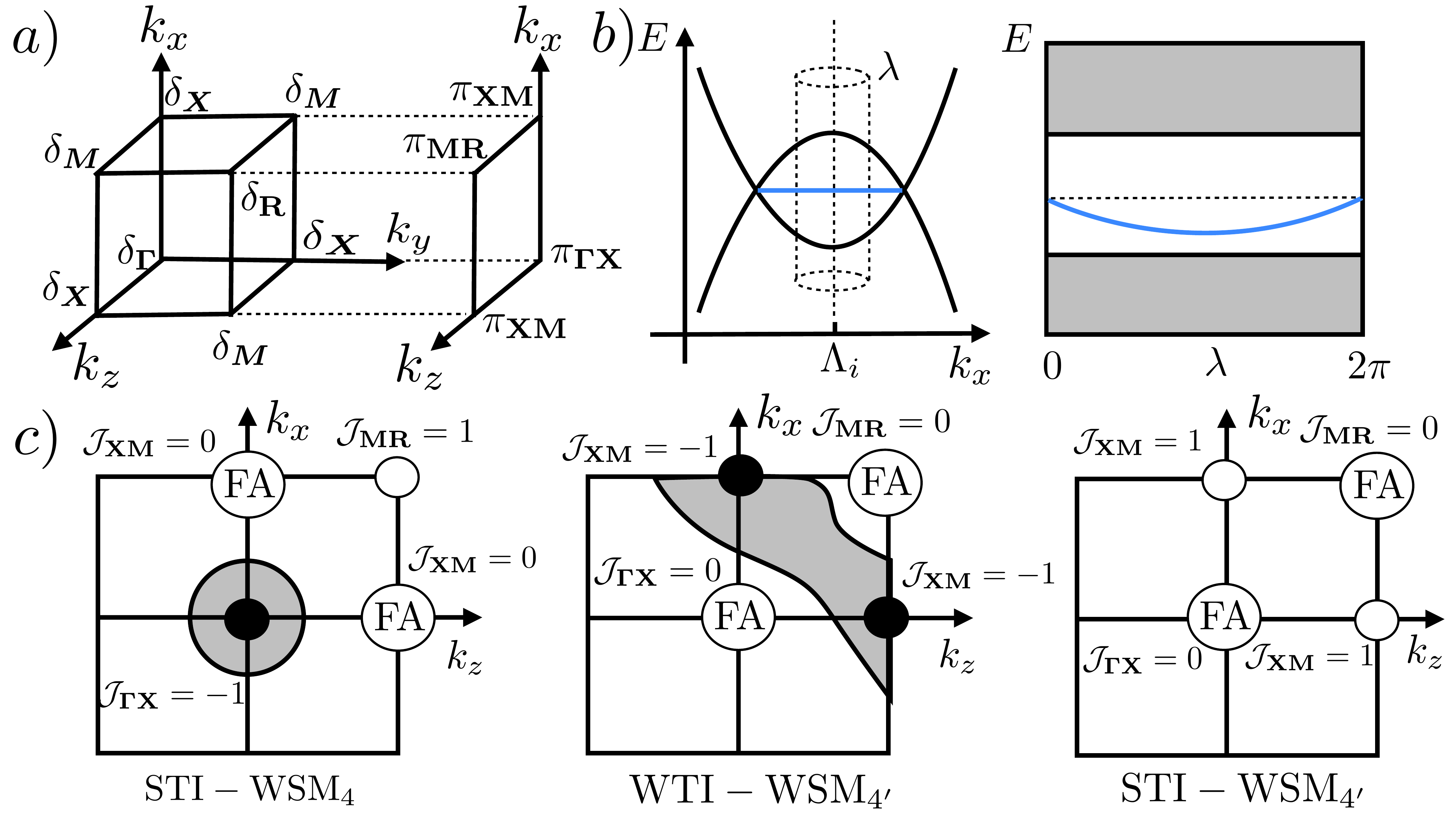}
\caption{\label{Fig:index} (Color online) Construction of the topological index:
a) Projection of the different bulk band parity products $\delta_{i}$ defining the time reversal polarization $\pi_{a}$ at each surface momenta $\Lambda_{a}$. The sign of all $\pi_{a}$ determine the presence (absence) of TI surface states \cite{FKM07,FK07}.
b) A cylindric 2D Hamiltonian around a bulk time reversal invariant momenta $\Lambda_i$ parametrized by $k_y$  and $\lambda \in [0,2\pi)$ (left panel). 
For open boundary conditions in $k_{y}$, the presence (absence) of a Fermi arc intersecting the cylinder, shows up as one (no) midgap state, defining $C_{a}=0 (1)$ (right panel). 
Lower panels: Schematic interface Brillouin zone including the value of Eq.~\eqref{eq:index} for three different domain wall configurations (from left to right): STI--WSM$_{4}$, WTI--WSM$_{4'}$ and STI--WSM$_{4'}$. The shaded regions enclose 2D surface massless Dirac states separated from the Fermi arcs (FA) in momentum space.
}
\end{figure}

To locate the TI Dirac surface states in momentum space we follow Refs.~\onlinecite{FKM07,FK07} and define for each bulk time reversal invariant momentum $\Lambda_i$ the product of the parity eigenvalues $\delta_{i}$ of the filled Kramers pairs. 
For a surface perpendicular to a lattice vector, each time-reversal invariant surface momenta $\Lambda_a$ is a projection of two bulk momenta $\Lambda_{i}$ and $\Lambda_{j}$, see Fig.~\ref{Fig:index}(a). 
It has associated with it the time reversal polarization $\pi_{a}=\delta_{i}\delta_{j}=\pm 1$ that determines the number and connectivity of the Dirac surface states~\cite{FKM07,FK07}. 
Namely, for any path connecting $\Lambda_{a}$ and $\Lambda_{a'}$ with $a \neq a'$ there is an odd (even) number of crossings at the Fermi level if $\pi_{a}\pi_{a'}=-1\,(1)$.
For $H_{\mathrm{TI}}$ in~\eqref{H} $\delta_{i}=-\mathrm{sign}[m_{i}]$. 

To similarly locate the Fermi arcs we employ the construction shown in Fig.~\ref{Fig:index}(b). 
Namely, we define a 2D Hamiltonian by restricting the 3D Hamiltonian~\eqref{H} to a cylinder enclosing a 3D time reversal invariant momenta $\Lambda_i$ but none of the Weyl nodes, and parameterize it by $(k_y,\lambda)$ with $\lambda \in [0,2\pi]$ describing circles in the $(k_x,k_z)$ plane.
If a Fermi arc exists between the surface Brillouin zone projections of two given Weyl nodes around $\Lambda_a$ it will cross this momenta as long as there is particle-hole symmetry~\cite{Ran06,JRL14}.
For open boundary conditions in $k_{y}$, the 2D Hamiltonian has a midgap state at the intersection of the cylinder and the Fermi arc, see Fig.~\ref{Fig:index}(b).
This enables us to define an index $C_{a}=0\,(1)$ that counts if there is a Fermi arc crossing $\Lambda_a$ (or not).
\footnote{We note that $C_{a}$ matches one minus the Chern number of a 2D Chern insulator Hamiltonian defined by a momentum space cut perpendicular to the line connecting the Weyl nodes if the WSM has only \emph{two} Weyl nodes.}

The two quantities $C_{a}$ and $\pi_{a}$ are separately obtained from the bulk of the WSM and the TI respectively.
Since we are interested in a domain wall we combine them in the index $\mathcal{J}_{a}=0,\pm 1$ introduced in 
\eqref{eq:index} associated to each interface momentum $\Lambda_{a}$.
From the index $\mathcal{J}_{a}$ we deduce the occurrence of nontrivial surface phenomena as follows.
At every $\Lambda_{a}$ where $\mathcal{J}_{a}=0$ a Fermi arc occurs.  
At a $\Lambda_{a}$ for which $\mathcal{J}_{a}\neq0$ the Fu-Kane criterion described above directly applies and analysis of the $\pi_{a}$ determines the existence of Dirac surface states. 
Hence, the key physical content captured by $\mathcal{J}_{a}$ is that both types of states at any given $\Lambda_a$ are mutually exclusive. 
We note that $C_{a}$, and by extension $\mathcal{J}_{a}$, relies on the fact that the Fermi arc crosses the 2D cylinder shown on the left panel of Fig.~\ref{Fig:index}(b), which is guaranteed even if $\mathcal{I}$ breaking terms are present, as long as there is no gap opening in the WSM and particle hole symmetry is respected. 
To exemplify the use of \eqref{eq:index} we apply the presented construction to the two domain walls considered above, as shown schematically in Fig.~\ref{Fig:index}(c).
First, for the STI--WSM$_{4}$ case we find that $C_{a}=0$ for $a =(k_{x},k_{y})=(0,\pi)$ and $(\pi,0)$, which are thus intersected by Fermi arcs.
Second, $C_{a}=1$ and $\pi_{a}=-1,1$ for $\Lambda_a =(0,0),(\pi,\pi)$ respectively, indicating an odd number of crossings at the Fermi level between those two surface momenta represented by a shaded circle in Fig.~\ref{Fig:index}(c) left panel. 
These conclusions are in perfect agreement with the numerical results shown in the upper panels of Fig.~\ref{Fig:interpol}.
For the second domain wall, the WTI--WSM$_{4'}$, the same procedure predicts two (an even number because of the WTI) Dirac surface states centered around $(0,\pi)$ and $(\pi,0)$ and two Fermi arcs crossing $(0,0)$ and $(\pi,\pi)$, see  Fig.~\ref{Fig:index}(c) centre panel.
Again, this agrees with the numerical results (see Fig.~\ref{Fig:interpol} lower panels).

An immediate consequence of our analysis is that not all WSM--TI interfaces host coexisting surface states, even when $\mathbf{b}$ is aligned perpendicular
to the domain wall. 
For instance, a domain wall involving the interpolation $(b_1, M_{1})\in$ STI $\to (b_2, M_{2}) \in$ WSM$_{4'}$ imposes, through $\mathcal{J}_{a}$,
that the Dirac nodes and the WSM Fermi arcs must cross $E=0$ at the same $\Lambda_{a}$, Fig.~\ref{Fig:index}(c) right panel. 
Since, as described above, Dirac states can occur only at $\Lambda_{a}$ where there are no Fermi arcs, 
only the two Fermi arcs corresponding to the WSM$_{4'}$ phase exist and cross $(k_{x},k_{z})=(0,0),(\pi,\pi)$. 
Consistent with our analysis based on $\mathcal{J}_{a}$, and as shown in the rightmost panels in Fig. \ref{Fig:interpol}, the inclusion of $b_{0}$ does not alter the coexistence of the surface states as long as it does not drive a phase transition to an insulator.

\emph{Optical conductivity of the surface states---}The presence of the surface states alters the response to an external electromagnetic field, which can be probed by optical spectroscopy.
The reflection coefficients determine the optical response and are related to the optical conductivity composed of a bulk and a surface state contribution~\cite{Born99}.
Experimentally, the bulk optical signature of TIs has been accessed via optical spectroscopy~\cite{LFP10} and the bulk WSM optical signature is theoretically well understood~\cite{HQ14,AC14}.  
Here, we compute the optical conductivity of the surface states reported above; in particular, the optical conductivity of the Fermi arc, to our knowledge, has not been calculated before.

We are interested in the linear response, long wavelength limit, where the incoming radiation has frequency $\omega$ and the momentum transfer satisfies $\mathbf{p}\to 0$.
The Fermi arc and the massless 2D Dirac fermion exist in separated parts of the Brillouin zone.
The total interface conductivity is therefore given by the sum of their individual and independent contributions $\sigma^{ij}_\text{surf}(\omega)=\sigma^{ij}_\text{df}(\omega)+\sigma^{ij}_\text{fa}(\omega)$.
The optical conductivity of a single 2D Dirac fermion, $\sigma_\text{df}^{xx}(\omega)=\sigma_\text{df}^{zz}(\omega)=\frac{\pi}{8}\frac{e^2}{h}$, is well known from the context of graphene~\cite{CGP09} and is isotropic and independent of the frequency $\omega$.
The Fermi arc can be modeled as a single chiral Fermion $\psi_{+}$ with definite chirality, chosen to be positive without loss of generality. Its Lagrangian is  
$\mathcal{L}= \psi^{\dagger}_{+}(i\partial_{0}+i\partial_{z})\psi_{+}$. 
To calculate its contribution we use the Kubo formula
\begin{equation}
	\mathrm{Re}\,\sigma^{ij}(\omega)=-\lim_{\mathbf{p}\to 0}\frac{1}{\omega}\mathrm{Im}\,\Pi^{ij}(\omega,\mathbf{p})
\end{equation}
that expresses the optical conductivity in terms of the polarization tensor $\Pi^{ij}(\omega,\mathbf{p})$.
We find (see~\footnote{See supplementary material for further details on the calculation of the optical conductivity of the Fermi arc.} for details of the calculation)
\begin{eqnarray}\label{eq:faopt}
\mathrm{Re}\,\sigma^{zz}_\text{fa}(\omega)&=&\dfrac{e^2\kappa_{0}}{2\pi^2\omega},\; \mathrm{Re}\,\sigma^{xz}_\text{fa}(\omega)=\mathrm{Re}\,\sigma^{xx}_\text{fa}(\omega)=0,
\end{eqnarray}
where $2\kappa_{0}$ is the separation between Weyl nodes in momentum-energy space. 
This prefactor reflects the fact that the Fermi arc only exists on a bounded part of the 2D surface Brillouin zone delimited by the surface Weyl node projection.
The optical conductivity of the Fermi arc is thus highly anisotropic and divergent as $\omega\rightarrow0$ in the clean limit. 

The total interface optical conductivity 
$\sigma^{ij}_\text{surf}(\omega)$ 
could be measured in a setup such as the one schematically shown in the lower left panel of Fig.~\ref{Fig:phasediag} where a TI thin film is deposited on top of a WSM, inspired by existing optical probes~\cite{LFP10}.
The TI bulk response vanishes for frequencies less than the bulk gap while the bulk WSM is proportional to $\omega$~\cite{HQ14,AC14}. Thus, for sufficiently low frequencies the response is determined by the surface and the coexistence can be probed by measuring the anisotropic Drude-like peak given in Eq. \eqref{eq:faopt} and a constant isotropic contribution from the 2D Dirac states.

\emph{Discussion and conclusions---}In this work, we have numerically demonstrated the possibility for Fermi arcs and 2D Dirac fermions to coexist at the
same surface, the interface of a Weyl semimetal and a topological insulator, in spite of explicit $\mathcal{T}$ symmetry breaking.
This is only possible if they do not coexist in the same region of reciprocal space and the time reversal breaking ${\bf b}$-vector of the WSM is perpendicular to the domain wall direction.
We have introduced a heuristic topological index $\mathcal{J}_{a}$, based on bulk topology, that can predict if and where the surface states are realized. This index captures the universal features of the bulk phases independent of the crystalline symmetry. Thus, it applies also to systems with
rhombohedral symmetry, (e.g. the Bi$_2$Se$_3$ family) that share the structure 
of the generic Hamiltonian studied here~\cite{LQZ10}. Even though Fermi arcs and Dirac cones can coexist, 
the latter are not as robust as those at TI--trivial insulator interface. 
A component of the ${\bf b}$-vector parallel to the domain wall direction
acts a Zeeman-term and gaps out the Dirac cone.  
The optical conductivity of the Fermi arc is found to be highly anisotropic and
therefore optical spectroscopy serves as a probe of coexistence of 2D
Dirac and Fermi arc surface states. 
In sum, our results uncover the interplay of distinct topological
bulk phenomena, topological insulators and semimetals, by showing that surface states with different nature can coexist at the same surface.

\emph{Acknowledgements---}A.G. G. is grateful to E. V. Castro for inspiring discussions from which this work originated. We thank F. de Juan and W. Witczak-Krempa for interesting comments and remarks, and T. Ojanen for discussions.

%

\newpage

\appendix

\begin{widetext}
\appendix

\section{\label{sec:app} Appendix: Optical conductivity of the surface states}
In this appendix we derive the linear response of the surface states to an external electromagnetic field in the long wavelength limit, where the incoming radiation has frequency $\omega$ and the momentum transfer satisfies $\mathbf{p}\to 0$.
Since the Fermi arc and the massless 2D Dirac fermion live
in distinct regions of the surface Brillouin zone we calculate separately their contributions to the conductivity, labelled
$\sigma^{ij}_\text{df}(\omega,\mathbf{p})$ and $\sigma^{ij}_\text{fa}(\omega,\mathbf{p})$ respectively. Mathematically
\begin{equation}\label{eqapp:current}
j^{i}(\omega,\mathbf{p})=\sigma^{ij}(\omega,\mathbf{p})E_{j}(\omega,\mathbf{p})
=\left[\sigma^{ij}_\text{df}(\omega,\mathbf{p})+\sigma^{ij}_\text{fa}(\omega,\mathbf{p})\right]E_{j}(\omega,\mathbf{p}),
\end{equation}
where the current density $j^{i}(\omega,\mathbf{p})$ and the external electric field $E_{i}(\omega,\mathbf{p})$ depend on the external frequency $\omega$ and momentum $\mathbf{p}$.\\

For completeness and to establish notation, we begin by reviewing the calculation of the contribution from the single massless 2D Dirac fermion. This can be adapted from known results in the context of graphene~\cite{CGP09} by simply dropping the valley and spin degeneracy factors. 
In linear response, we express the current in terms of the vector potential $A_{j}$ as $j^{i}=\Pi^{ij}A_{j}$,	
written in terms of the spatial component of the polarization tensor $\Pi^{\mu\nu}$.  To first order, this propagator is given by the bubble diagram shown in Fig. \ref{Fig:bubble} that represents the integral
\begin{equation}\label{eqapp:bubbleint}
i\Pi^{\mu\nu}_\text{df}(p)=-(-ie)^2\mathrm{Tr}\int \dfrac{d^3k}{(2\pi)^3} G(k)\gamma^{\mu}G(k+p)\gamma^{\nu},
\end{equation}
with $G(k)=i/\slashed{k}$, $k^{\mu}=(k_{0},\mathbf{k})$, and $p^{\mu}=(\omega,\mathbf{p})$. We use Feynman's
slashed notation defined through $\slashed{k}=k_{\mu}\gamma^{\mu}$. 
The integral, written as
\begin{equation}\label{eqapp:bubble}
\Pi^{\mu\nu}_\text{df}(p)=(-ie)^2\int \dfrac{d^3k}{(2\pi)^3}\dfrac{\mathrm{Tr}\left[\slashed{k}\gamma^{\mu}(\slashed{k}+\slashed{p})\gamma^{\nu}\right]}{k^2(k+p)^2},
\end{equation}
has the precise same form as the QED(2+1) bubble and it is linearly divergent 
by naive power counting. Nonetheless, the final result is finite and can 
be obtained by using for example dimensional regularization.
By analytic continuation to dimension $d$ the result can be written in 
terms of the auxiliary Feynman parameter $x\in[0,1]$ as~\cite{Peskin} 
\begin{equation}\label{eqapp:dbubble}
i\Pi^{\mu\nu}_\text{df}(p)=-i(p^2 g^{\mu\nu}-p^{\mu}p^{\nu})\dfrac{2e^2}{(4\pi)^{d/2}}\mathrm{Tr}\left[\mathbbm{1}\right]\int_{0}^{1} dx\dfrac{\hspace{2mm} x(1-x) \Gamma(2-d/2)}{(-x(1-x)p^2)^{2-d/2}},
\end{equation}
where $\mathrm{Tr}\left[\mathbbm{1}\right]$ is the trace over the dimension of the Pauli matrices. This
last result will prove useful for the computation of the Fermi arc contribution to the conductivity.
For the 2D massless Dirac cone one can set the spacetime dimension to $d=3$ 
and $\mathrm{Tr}\left[\mathbbm{1}\right]=2$ to evaluate the integral and obtain \cite{GGV94}
\begin{equation}\label{eqapp:dbubbleres}
i\Pi^{\mu\nu}_\text{df}(p)=-i(p^2 g^{\mu\nu}-p^{\mu}p^{\nu})\dfrac{4e^2}{(4\pi)^{3/2}}
\dfrac{-i\pi^{3/2}}{8\sqrt{\omega^2-\mathbf{p}^2}}=-\dfrac{e^2}{16}\dfrac{(p^2 g^{\mu\nu}-p^{\mu}p^{\nu})}{\sqrt{\omega^2-\mathbf{p}^2}}.
\end{equation}
We are interested in the real part of the optical conductivity, which can for example be probed by optical spectroscopy, related to the polarization tensor $\Pi^{ij}$ through
\begin{eqnarray}\label{eqapp:Kubo1}
\mathrm{Re}\,\sigma^{ij}(\omega,\mathbf{p})&=&-\lim_{\mathbf{p}\to 0}\dfrac{1}{\omega}\mathrm{Im}\,\Pi^{ij}(\omega,\mathbf{p}).
\end{eqnarray}
Using that $\partial_{\mu} j^{\mu}=0$  
\begin{equation}
\mathrm{Re}\,\sigma^{ij}(\omega,\mathbf{p})=-\lim_{\mathbf{p}\to 0}\dfrac{\omega}{p_{i}p_{j}}\mathrm{Im}\,\Pi^{00}(\omega,\mathbf{p})
=\lim_{\mathbf{p}\to 0}\dfrac{\omega}{p_{i}p_{j}}\mathrm{Im}\left[\dfrac{e^2}{16}\dfrac{\mathbf{p}^2}{\sqrt{v_{F}\mathbf{p}^2-\omega^2}}\right]
\label{eqapp:final2D}
=\dfrac{\pi}{8}\dfrac{e^2}{h}\delta^{ij}.
\end{equation}
where in the last step we have introduced the Kronecker delta function $\delta^{ij}$ and restored $e^2/\hbar$ to obtain the optical conductivity for 2D massless Dirac fermions used in the main text~\cite{CGP09}.
\begin{figure}
\includegraphics[scale=0.085]{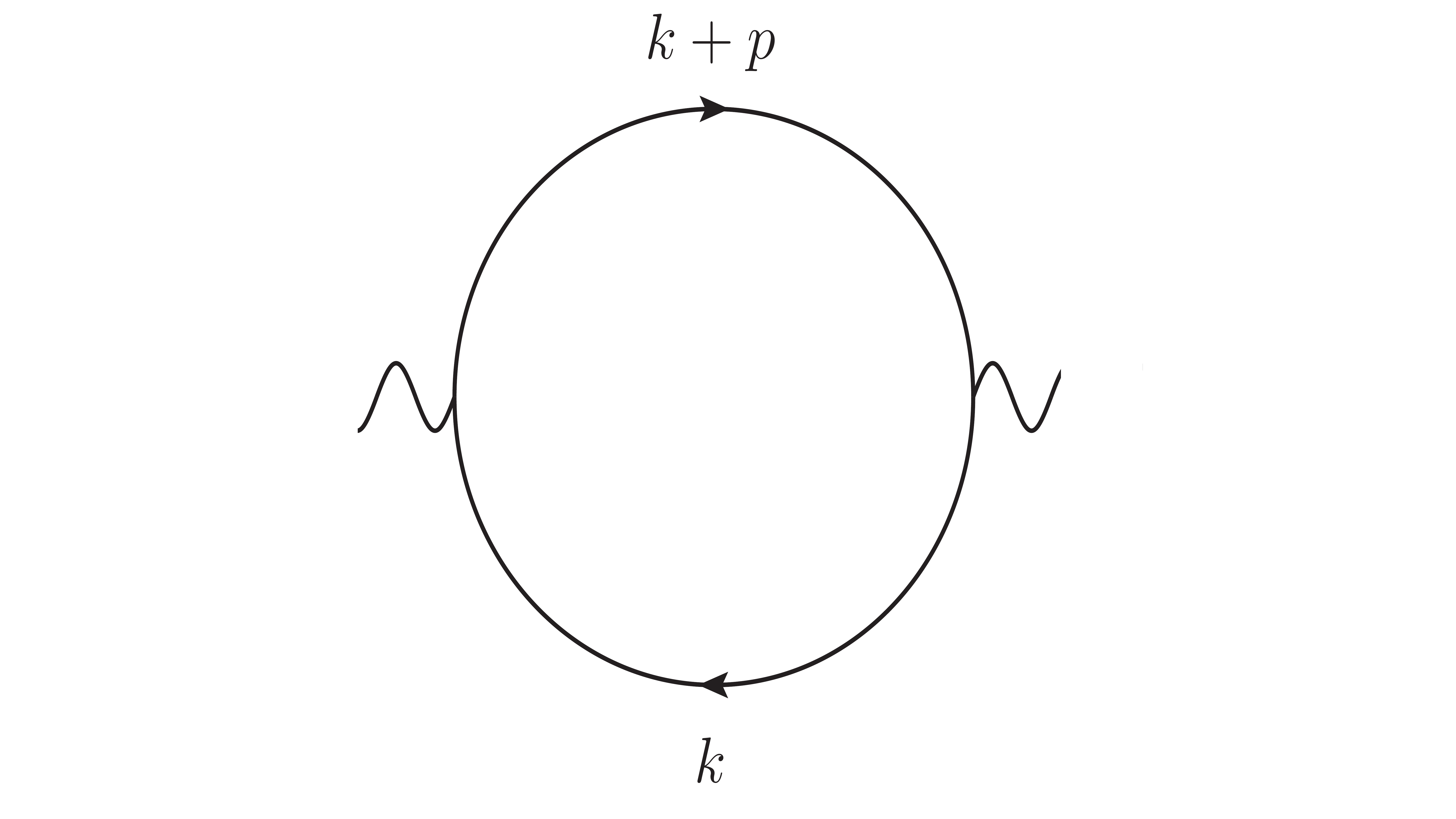}
\caption{\label{Fig:bubble} Polarization bubble representing the integral \eqref{eqapp:bubbleint}.
The solid curved lines represent a fermionic propagator $G_{k}$ evaluated at the corresponding momenta. For the 2D Dirac fermion it can be written as $G_{k}=i/\slashed{k}$ with $k^{\mu}=(k_{0},k_{1},k_{2})$ ($\slashed{k}=k_{\mu}\gamma^{\mu}$) while for the chiral Fermion describing the Fermi arc is given $G_{k}=i\mathcal{P_{+}}\slashed{k}/k^2$ with $k=(k_{0},k_1)$. The  wavy lines represent the external electromagnetic field carrying external momentum $p=(\omega,\mathbf{p})$.
}
\end{figure}

Now we want to calculate the contribution from the Fermi arc to the optical conductivity. The Fermi arc is a single chiral
Fermion and thus its Lagrangian can be written as
\begin{equation}
\mathcal{L}= \psi^{\dagger}_{+}(i\partial_{0}+i\partial_{1})\psi_{+}=\bar{\psi}\mathcal{P}_{+}\gamma_{\mu}\partial^{\mu}\psi,
\end{equation}
where in the last step we have written the chiral fermion in terms of the projection of a Dirac fermion with the help
of the projector operator $\mathcal{P}_{\pm}=\frac{1}{2}(1\pm\gamma_{5})$. We note that the $\gamma$ matrices
here do not correspond to the $\gamma$ matrices of the 2D calculation but instead can be chosen to have the representation$\gamma^{0}=\sigma_{y},\gamma^{1}=i\sigma_{x}$ and $\gamma^{5}=\gamma^{0} \gamma^{1}=\sigma_{z}$.

The propagator for such a state is $G_{k}=i\mathcal{P_{+}}\slashed{k}/k^2$ with $k=(k_{0},k_1)$. Note that the chiral fermion is different from the usual Dirac fermion in $1+1$ dimensions in that i) the propagator of the latter can be written as $G^{1+1}_{k}=i\mathcal{P_{+}}\slashed{k}/k^2+i\mathcal{P_{-}}\slashed{k}/k^2$ and ii) the former is embedded in a 2+1 dimensional space time, unlike its $1+1$ dimensional Dirac counter part.
For the calculation of the polarization tensor \eqref{eqapp:bubble} of the Fermi arc, the integral in one of the directions is constrained by the separation between Weyl points in momentum space $2\kappa_{0}$, a fact taken into
account by setting the limits of the first integral in $k_{x}$ to $-\kappa_{0}<k_{x}<\kappa_{0}$.
Therefore, for the Fermi arc case Eq. \eqref{eqapp:bubble} reads
\begin{equation}\label{eqapp:bubblefa1}
i\Pi^{\mu\nu}_\text{fa}(p)=
(-ie)^2\int^{\kappa_{0}}_{-\kappa_{0}} \dfrac{dk_{x}}{2\pi}\int\dfrac{d^2k}{(2\pi)^2}\dfrac{\mathrm{Tr}\left[\mathcal{P_{+}}\slashed{k}\gamma^{\mu}\mathcal{P_{+}}(\slashed{k}+\slashed{p})\gamma^{\nu}\right]}{k^2(k+p)^2}.
\end{equation}
Using the definition of $\mathcal{P_{+}}$ and the properties of the trace, the integral can be rewritten as:
\begin{equation}\label{eqapp:bubblefa2} 
i\Pi^{\mu\nu}_\text{fa}(p)=
\dfrac{(-ie)^2}{2}\dfrac{2\kappa_{0}}{2\pi}\int\dfrac{d^2k}{(2\pi)^2}\left[\dfrac{\mathrm{Tr}\left[\slashed{k}\gamma^{\mu}(\slashed{k}+\slashed{p})\gamma^{\nu}\right]}{k^2(k+p)^2}\right. 
+\left.\dfrac{\mathrm{Tr}\left[\gamma_{5}\slashed{k}\gamma^{\mu}(\slashed{k}+\slashed{p})\gamma^{\nu}\right]}{k^2(k+p)^2}\right].
\end{equation}
We can use the fact that $\gamma_{5}\gamma^{\mu}=-\epsilon^{\mu\nu}\gamma_{\nu}$ to write the second integral in the same
mathematical form as the
first by pulling the $\epsilon^{\mu\nu}$ tensor out. 
Then, we notice that both integrals are formally the same as \eqref{eqapp:bubble} but in one dimension less.
Thus, we can use \eqref{eqapp:dbubble} with $d=2$ and $\mathrm{Tr}[\mathbbm{1}]=2$ to write the final solution
\begin{equation}\label{eqapp:bubblefaf}
i\Pi^{\mu\nu}_\text{fa}(p)=\dfrac{ie^2 \kappa_{0}}{2\pi^2p^2}\left[(p^2 g^{\mu\nu}-p^{\mu}p^{\nu})\right. +
\left.(p^2 \epsilon^{\mu\nu}-\epsilon^{\mu\rho}p_{\rho}p^{\nu})\right].
\end{equation}
To relate this result with the actual surface conductivity notice that here $\mu,\nu \in \{0,1\}$
while for the 2D Dirac cone we had $\mu,\nu \in \{0,1,2\}$. This means that the Fermi arc has only one non-vanishing 
longitudinal component of the optical conductivity (i.e. $\sigma^{zz}$), while the 2D Dirac fermion has both $\sigma^{zz}$ and $\sigma^{xx}$ 
contributions. Using \eqref{eqapp:Kubo1}, the optical conductivity of the Fermi arc is therefore
\begin{subequations}
\begin{eqnarray}\label{eqapp:finalfermiarc}
\mathrm{Re}\,\sigma^{zz}_\text{fa}(\omega)&=&\dfrac{e^2\kappa_{0}}{2\pi^2\omega},\\
\mathrm{Re}\,\sigma^{xx}_\text{fa}(\omega)&=&0,\\
\mathrm{Re}\,\sigma^{xz}_\text{fa}(\omega)&=&0.
\end{eqnarray}
\end{subequations}
The total surface conductivity is given given by the sum of \eqref{eqapp:final2D} and
\eqref{eqapp:finalfermiarc} for $\sigma^{xx}$ and by \eqref{eqapp:final2D} for $\sigma^{zz}$.\\

\end{widetext}

\end{document}